\def\year{2019}\relax
\documentclass[letterpaper]{article} 
\usepackage{aaai19}  
\usepackage{times}  
\usepackage{helvet} 
\usepackage{courier}  
\usepackage[hyphens]{url}  
\usepackage{graphicx} 
\usepackage{graphicx}  
\usepackage{enumitem}
\title{Developing Computational Models of Social Assistance to Guide Socially Assistive Robots}
\author{Jason R. Wilson, Seongsik Kim, Ulyana Kurylo, Joseph Cummings, Eshan Tarneja\\
Northwestern University\\ 
2233 Tech Drive \\
Evanston, IL 60208 \\
jrw@northwestern.edu,\{josephkim,uk,jcummings,eshan\}@u.northwestern.edu
}

\begin{document}

\maketitle

\begin{abstract}
While there are many examples in which robots provide social assistance,
a lack of theory on how the robots should decide how to assist impedes progress in realizing these technologies.
To address this deficiency, 
we propose a pair of computational models to guide a robot as it provides social assistance.
The model of \textit{social autonomy} helps a robot select an appropriate assistance that will help with the task at hand while also maintaining the autonomy of the person being assisted.
The model of \textit{social alliance}  describes how a to determine whether the robot and the person being assisted are cooperatively working towards the same goal.
Each of these models are rooted in social reasoning between people, and we describe here our ongoing work to adapt this social reasoning to human-robot interactions.
\end{abstract}

\noindent 

%
Socially assistive robots (SARs) provide social assistance instead of physically intervening.
For many SARs, assistance comes in the form of verbal cues and simple gestures (e.g., \cite{mataric2007socially,wilson2018supporting}).
Many approaches are used to guide the robot in determining what to say and which gestures to use, but the reasoning typically focuses on the task and not the social relationship between the robot and the user - the person being assisted.
We propose that our ongoing work in developing computational models of social assistance will provide a theoretical foundation for informing how a robot provides social assistance.

In the work we present here, we draw upon research in therapeutic settings, in which a person is responsible for providing therapeutic assistance to another person.
We believe that how people socially interact with each other while giving and receiving assistance can inform how a robot can behave when giving assistance to a person.
In many therapeutic settings, a clinician attempts to build a \textit{working alliance} with a client by building rapport and ensuring that their therapeutic goals align \cite{tickle2002client}.
Additionally, it is important that a clinician assist a client in a manner that supports the \textit{dignity} and \textit{autonomy} of the person by providing assistance in accordance with the needs of the person \cite{wilson2016autonomy}. We adapt the work in clinical practices and extend prior research in socially assistive robots to support the development of \textit{computational models of social assistance}.  To this end, we propose the following goals:


\begin{enumerate}
	\item Develop a model of \textit{social autonomy} that enables an autonomous robot to provide assistance in accordance with the needs and preferences of the user. 
	\item Develop a model of \textit{social alliance} that is based on human models of building a working alliance and describes the developmenet of a social alliance through 
	an alignment of the robot's and user's goals.
\end{enumerate}	

We describe here our current progress towards developing these models. We describe each model, the current state of each model, next steps in developing the model, and proposed hypotheses and evaluation plans.

%




\section{Example Domain}

As a demonstration of how social assistance models would be applied, we focus on one application in which a social robot helps a person with sorting medications.
In this task, a person places pills on a sorting grid such that all the constraints of the prescriptions are met while adjusting for scheduling constraints imposed by life events (e.g., a doctor's appointment requiring a medication to be taken earlier than usual) \cite{Wilson2016des}.
During the task, a robot provides social assistance in the form of encouragements, questions, suggestions, and gestures.  For example, the robot may say ``Good job'' 
or ``There are too many blue pills on Tuesday morning''. 

\section{Model of Social Autonomy}
A model of social autonomy, which includes domain, assistance, and need models, is used to ensure that a robot is able to provide assistance that supports the autonomy of the user.
We define autonomy as being able to freely make choices and act in accordance with one's goals, needs, and desires. 
Autonomy can be inhibited by preventing a person from doing what he or she wants to do \cite{Nordenfelt2004}.
Conversely, people can feel more autonomous when they are able to choose a course of action \cite{hertz2002relationships}, be in control \cite{hertz2002relationships,weinstein2012index},
and have the necessary resources to take desired actions
\cite{hertz2002relationships}.

\begin{figure}
    \centering
    \includegraphics[width=\columnwidth]{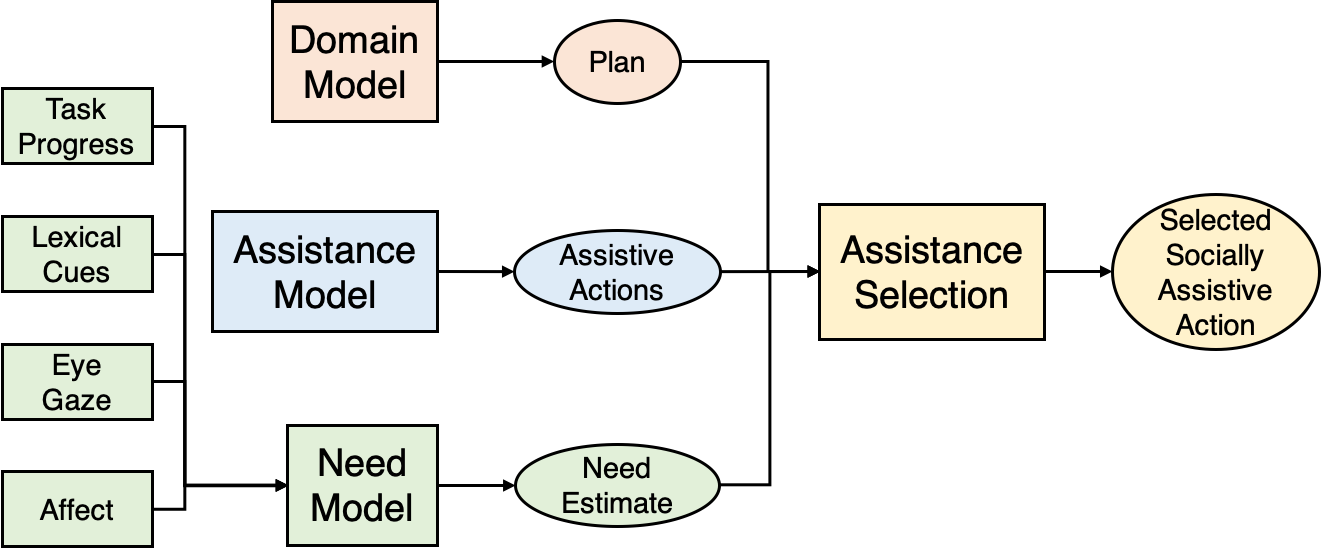}
    \caption{Models of need and assistance are used with a model of the domain to select a socially assistive action to help with the task while maintaining the autonomy of the assisted.}
    \label{fig:auto}
\end{figure}

The model of social autonomy is based on a principle that we propose stating that
the autonomy of the person being assisted is inversely proportional to the difference between the amount of assistance provided and the amount need.
Towards this principle, we are developing a set of models that compose the model of social autonomy.
Figure~\ref{fig:auto} shows how the domain, assistance, and need models combine to provide information for the assistance selection algorithm.
The domain model allows the robot to understand the task at hand, and
The assistance and need models capture some of the social understanding necessary in determining how and when to assist.
The results of these models are used by an assistance selection algorithm that selects a \textit{socially assistive action}, an action to be performed by the robot to assist the user in completing a task.


\subsection{Domain Model}

\begin{figure}
    \centering
    \includegraphics[width=\columnwidth]{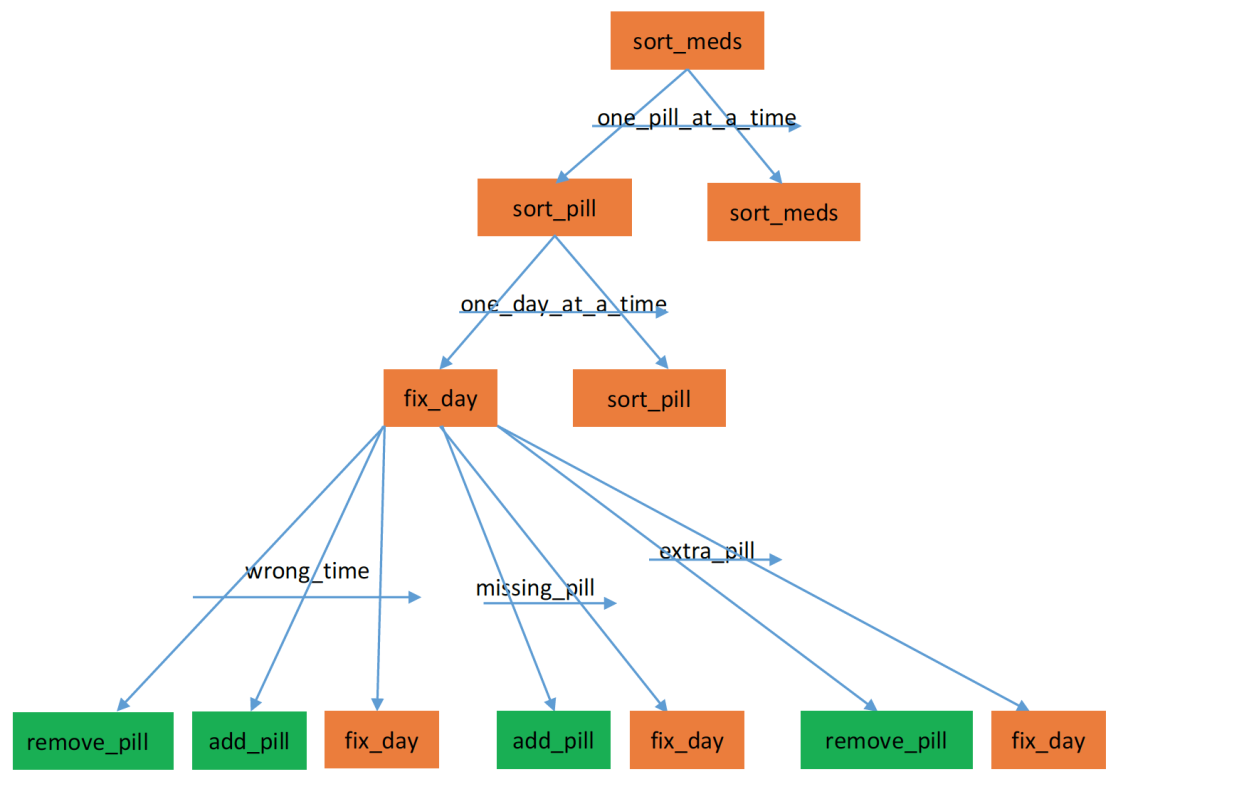}
    \caption{HTN used to build a plan for sorting medications.}
    \label{fig:htn}
\end{figure}

To describe the domain of the task with which a robot is to assist, we use a hierarchical task network (HTN) (see Fig.~\ref{fig:htn}).
We use pyhop, a python implementation of the SHOP algorithm, to generate a plan from the HTN \cite{nau2013game}.  
The plan consists of  a sequence of actions to add and remove pills to/from the sorting grid in order to complete a medication sorting task.
The first action in the plan represents the next step in the task with which the robot will attempt to provide assistance.

\subsection{Assistance Model}

To represent how much assistance is provided with an action that a robot performs, we define a set of \textit{assistive actions}.  Each social assistance is associated with an action that the user may perform  to complete a task (e.g., \texttt{remove\_pill} and \texttt{add\_pill}) and a level of assistance \cite{wilson2018needbased}.
A level of assistance may be from one to three, representing the first three levels of assistance described in the Performance Assessment for Self-care Skills Manual \cite{RogersHolm1994}. 
These three levels of assistance represent purely social actions, whereas the higher levels require some form of physical action.

\subsection{Need Model}

In order to provide assistance in accordance with the amount of need for assistance, a robot must be able to calculate a \textit{need estimate}, a perception of how much need for assistance the user has.
We combine many indicators to estimate need, including whether the person is making progress in a task, verbal requests for help, the direction in which the person is looking, and the emotions being expressed by the user.  We describe each of these indicators, how they are used to estimate how much need for assistance a person has, and ongoing work to combine the indicators to create a holistic perspective on the user.


\paragraph{Task Progress}
To determine if a person needs assistance in a task, a robot may examine the person's progress in the task.  Making a mistake or simply  not doing anything can be indicators that a person may need some assistance.
Conversely, if a person takes steps that bring the person closer to completing the task, then the person probably does not need assistance.
One approach to measuring progress in a task examines whether the length of a plan to complete the task changes \cite{wilson2018needbased}. If the number of actions remaining decreases, then progress is being made and assistance may not be needed. On the other hand, if the plan length remains the same or increases, then the person could need some assistance.
Since a lack of progress can also indicate that a person is contemplating or considering alternative approaches, it cannot be automatically assumed that a person immediately needs assistance.  
Instead, multiple cues are necessary to determine if the person actually needs assistance, and we describe next a set of social cues to recognition when and how much assistance a person needs.

\paragraph{Lexical cues}
There are some things a person can say that make it clear a person needs assistance.
For example, a person might say something like ``Please help me'' or ``I don't understand''.
Other utterances are more complicated and may require a semantic understanding. 
For example, a person could say, ``I forget when my doctor's appointment is,'' which can be interpreted as asking the robot to remind the person when the appointment is.
Our models thus far has used keyword matching to indicate when a person needs assistance based on what the person says \cite{wilson2018supporting,wilson2018needbased}.



\paragraph{Eye Gaze}
Eye gaze is a subtle communication cue which most people naturally engage in. It is used to moderate conversation and provide feedback in social interactions \cite{kleinke1986gaze} and can thus be informative for gauging when someone is struggling with a task or needs help.
When engaged in a task, a person will be actively looking at the task itself. With this consistent gaze, the person is actively working on the task and thus may not require any assistance. If the person looks away, the person may be contemplating and does not necessarily need assistance.

However, when the person looks away from the task to look at the robot, the person might be seeking some assistance or feedback from the robot. Mutual gaze, where the person and the robot look at each other, is natural in social interactions \cite{admoni2017social}, and we have found that when a person initiates a mutual gaze with the robot that a person is trying to communicate that assistance is needed from the robot  \cite{kurylo19using}.

A similar gaze pattern is confirmatory gaze, in which a person cycles between glances at the robot and glances at the task. This pattern is often an indicator of need for assistance; people engage in this pattern right after they have taken a step in the task and want to confirmation that they are on the right track. As with mutual gaze, a confirmatory gaze pattern has been shown to be a useful indicator in predicting when a person needs assistance \cite{kurylo19using}.


\paragraph{Emotion}

Emotions can be a useful indicator as to whether a person needs assistance, and there are many ways in which an emotion can be recognized in a person (e.g., face, speech, body posture).
We start by looking at the textual content of what is said in a conversation, building upon recent work in recognizing emotions in short conversations - which provide more context than a single utterance. 
Using a large data set of short Twitter conversations \cite{chatterjee-etal-2019-semeval}, we developed a Bayesian model to categorize then end of the conversation as being one of three emotions: happiness, sadness, or anger \cite{cummings-wilson-2019-clark}. 
Our model performed on par with a baseline deep learning model while requiring 1/6 the data and 1/70 of the time to train.
We believe that this model can be extended to incorporate additional emotions, namely boredom, challenge, and frustration.
These emotions, along with anger, are likely to be good indicators that a person needs assistance.



\paragraph{Combining Indicators}
Each of the approaches described above independently infers whether a person needs assistance, and we are currently working on approaches to combine these indicators into a single estimate of how much need the person has.
We view this has a decision fusion problem \cite{atrey2010multimodal}, for which we are experimenting with a linear weighted fusion approach.  Some indicators, such as task progress, are reliable indicators of when a person needs assistance and would be weighted higher.  
Additionally, the significance of each indicator can be learned, by learning the weights or using other learned approaches, such as Bayesian inference.



\subsection{Assistance Selection}
To select the socially assistive action, we use the Hint Engine, which selects an appropriate assistance by taking the first action in a plan for the remainder of the task, looks up a list of assistive actions for that user action, then selects the assistance matching the amount of need \cite{wilson2018needbased}.
By selecting the assistive action that matches the amount of need, a robot is able to provide the right amount of assistance such that the autonomy of the user is maintained.

The approach used in the Hint Engine assumes that any time a user action is required, the same types of assistive actions are applicable.
However, context may vary, and the reason for doing a task can influence the type of assistance that would be applicable.
As a result, we have updated the algorithm to incorporate some justification for a task by incorporating information from an HTN used to construct the plan.
For each user action in the resulting plan, we record the tasks and methods that led the planner to specifying a given action.
For example, in the HTN in Figure 1, an \texttt{add\_pill} action can be included in the plan for the methods \texttt{wrong\_time} and \texttt{missing\_pill}, and which method is used to create the plan can be used to justify the action.
Given a list of justifications for an action, the algorithm to select an assistance that matches the amount of need is updated to select the justification relative to the amount of need.  Then, we retrieve an assistance corresponding with the given justification.  Thus, if the \texttt{add\_pill} action needs to be done next, the robot can provide an assistance suggesting this action while providing some explanation.  In other words, the robot can say 
``You have a pill at the wrong time on Tuesday'' or ``You are missing a pill on Tuesday'' for the action \texttt{add\_pill} depending on which reason a pill needs to be added.

\subsection{Proposed Hypothesis and Model Validation}

Our hypothesis is that when assistance is provided in accordance with the amount of need a person has, then autonomy can be maintained.
Conversely, if assistance and need do not match (assistance is too little or too much) then autonomy is inhibited.
More formally, we register following hypothesis to be validated in future work:

\begin{itemize}[leftmargin=9mm]
    \item[(H1)]  A person being assisted in a task will feel more autonomous if the assistance provided matches the amount of need for assistance.
\end{itemize}

This hypothesis assumes that we can measure and infer how much need for assistance a person has.  To this end, we have validated that models of mutual gaze and confirmatory gaze are good indicators of when a person needs assistance \cite{kurylo19using}, and we are in progress on validating models for task progress, lexical cues, and emotion.

Our hypothesis also requires us to be able to knowing how much autonomy a person being assisted has, we need a measure of autonomy. Existing measures such as the Inventory of Autonomous Functioning are not applicable as they capture trait autonomy \cite{weinstein2012index}. Instead, we need to measure a transient sense of autonomy that is in direct response to the receiving assistance on a task. 
As a result, we are developing a new measure of autonomy, which is currently being validated in experiments in which a person receives assistance and answers a survey of 94 items that reflect issues of autonomy (i.e., authorship,  interest-taking, and susceptibility to control).



\section{Model of Social Alliance}

A model of social alliance, which includes recognizing the goals and intentions of the user, 
is used to ensure that a robot is able to build a cooperative alliance with the human user(s) in which the goals of the user(s) are aligned with the goals of the robot.
In a therapeutic setting, a working alliance is formed as individuals collaborate with one
another to develop common goals and a shared sense of responsibility \cite{bordin1979generalizability}.
Similarly, a robot needs to be working towards the same goal as the user, 
but it might not always be clear if the robot and the person are on the same page.
The robot may need to make observations of the user's behaviors to determine the goals and intentions of the user.
We describe here our plans for future work to build a model of social alliance, which encompasses the social reasoning necessary to develop a working alliance.

To infer the goals and intentions of the user, we propose a computational model using a combination of plan-based goal recognition and Theory of Mind (ToM).
Plan-based goal recognition can be used to determine whether observed actions align with a plan for a given goal \cite{holler2018plan}.
Goal recognition relies on a shared model of the domain, which may be the case between virtual agents.  

Additionally, we consider a perhaps more subtle goal distinction - determining  whether the user intends to cooperate or not.  
Recent ToM models have shown to be effective in recognizing whether an agent is intending on cooperating with another agent.
For example, a Bayesian ToM model can effectively recognize the desires and beliefs of an agent by finding the most probable objective given the trajectory of the agent's movements \cite{baker2011bayesian}.
Since a Bayesian approach may require a significant amount of training data, we consider an alternative approach that uses analogical inference.
AToM is an analogical ToM model that compares a given situation with a prior example, if the situations are similar enough, AToM will use the prior example to infer the other agent's internal states, including its goals and intentions \cite{rabkina2019analogical}.
By using analogical inference, AToM is able to make accurate inferences with only a few prior examples.


We propose combining a goal recognizer and a ToM model, using a goal recognizer to identify the user's goals when the shared models are sufficient and supplementing it with a ToM model to recognize more subtle desires and intentions.
For example, consider a scenario in which a robot is given a goal to help the user sort all the pills of a given medication to be taken in the morning.  However, the person places all the pill in the afternoon instead. 
If the goal recognizer detects that the user has executed a different goal, 
then the robot needs to take actions to fix this different.  If the user is executing a permissible goal, then the robot may be able to adjust its own goals.  Otherwise, the robot needs to communicate that it has detected a difference and the user may need to adjust.
Alternatively, if the person does not want to sort his or her medications, the person may take no actions at all. If the robot has a prior example of the person not taking actions because the person does not intend to do the task, then AToM can be used to infer that the person has similar intentions in the current task.

\subsection{Proposed Hypothesis and Model Validation}




Since we propose  that a robot can build a social alliance by ensuring that the goals of the robot and the user align, we formulate the following hypothesis to be used to validate our model:
\begin{itemize}[leftmargin=9mm]
    \item[(H1)] Goal alignment positively contributes to the building of a social alliance.
\end{itemize}



To measure the dependent variable, social alliance, we will use a standard measure used in clinical settings, the Working Alliance Inventory \cite{horvath1989development}. In addition to being used in clinical practice, the Working Alliance Inventory uses in HRI research include 
measuring fluency \cite{hoffman2019evaluating} and evaluating a relational robot \cite{bickmore2010relational}. 

While building a social alliance is the ultimate goal of this model, it requires that the robot is able to infer the goals of the user.  As a result, we also need to evaluate how well our model is able to make this inference.  As such, we have the following sub-hypotheses:
\begin{itemize}[leftmargin=10mm]
    \item[(H1a)] Inferring a user's goal can indicate whether the goals of the user and the robot align.
    \item[(H1b)] Whether a goal of the user is to perform a task or not may be inferred from the user's actions.
\end{itemize}

\section{Conclusion}
These models are not divorced from each other but have many important connections.
If the goals of the robot and the user do not align, it is highly unlikely that the assistance the robot gives will be useful or appreciated.
Also, as a robot is giving assistance, it can continue to monitor whether their goals remain aligned.
If the person begins to need more assistance, this may be the result of the person no longer being interested in the task.
How the robot responds in this case can vary greatly, depending on the domain and the overarching goals of the robot.
In some cases, it may be applicable for the robot to complete the task for the person.
While in other cases, perhaps for a tutoring robot, the robot should not complete the task.  
Ultimately, the robot needs to be working towards the same goal as the user, and when the goals align, assistance has to be in accordance with how much need for assistance the person has.
The models of social assistance proposed here provide a theoretical foundation to guide a robot in determining if the robot's assistance is contributing to building an alliance and maintaining the autonomy of the user.



\bibliographystyle{aaai}
\bibliography{references}

\end{document}